\documentclass[prl,aps,twocolumn,superscriptaddress,nofootinbib,longbibliography]{revtex4-2}
\usepackage{graphicx} % Include figure files
\usepackage{dcolumn}  % Align table columns on decimal point
\usepackage{bm}       % bold math
\usepackage{bbm}       % bold math
\usepackage{amsmath}
\usepackage{epsfig}
\usepackage{color}
\usepackage{ulem}
\usepackage[colorlinks,linkcolor=black,citecolor=blue,urlcolor=blue,hyperindex,driverfallback=dvipdfm]{hyperref}
\usepackage{orcidlink}

\begin{document}
\title{Thermal Radiation Force and Torque on Moving Nanostructures with Anisotropic Optical Response}

\author{Juan R. Deop-Ruano\,\orcidlink{0000-0002-2613-5873}}
\email{juan.deop@csic.es}
\affiliation{Instituto de Qu\'imica F\'isica Blas Cabrera (IQF), CSIC, 28006 Madrid, Spain}
\author{Alejandro Manjavacas\,\orcidlink{0000-0002-2379-1242}}
\email{a.manjavacas@csic.es}
\affiliation{Instituto de Qu\'imica F\'isica Blas Cabrera (IQF), CSIC, 28006 Madrid, Spain}

\date{\today}

\begin{abstract}
Nanoscale objects moving relative to a thermal radiation bath experience a drag force due to the imbalance in their interaction with the blue- and redshifted components of the electromagnetic field. Here, we show that, in addition to this drag force, moving nanostructures with an anisotropic optical response experience a lateral force and a torque that substantially modify their trajectory. These phenomena emerge from the additional coupling between the electromagnetic field components polarized parallel and perpendicular to the trajectory, enabled by the anisotropic response of the nanostructure. This work unveils the intricate dynamics of anisotropic nanostructures moving in a thermal radiation bath.
\end{abstract}
\maketitle

The thermal and quantum fluctuations of the electromagnetic field give rise to counterintuitive forces and torques acting on neutral objects, commonly known as Casimir interactions \cite{C1948, EGJ09, L07, LMR10, RCJ11, DMR11, ECH15, LZ16, WDT16, MCK21, XJG22}. Generally, the effect of these interactions is to displace and reorient the objects to a configuration of minimal energy. For instance, the Casimir interaction between two conductive plates in vacuum results in an attractive force  \cite{C1948}, whereas for two objects with anisotropic response, a Casimir torque acts to align their principal axes \cite{PW1972, B1978, GGL15, SGP18, ACG20}. 
Beyond their fundamental importance, Casimir interactions are being explored for technological applications \cite{PSS20}. These include, among others, the implementation of self-assembled nanocavities \cite{MCK21, KKC24} and the passive trapping and manipulation of nanoscale objects \cite{ZLY19,SCD23}.

When two or more objects move relative to each other, Casimir interactions also constitute a source of friction that opposes their motion \cite{RIB22}. These frictional phenomena have been extensively studied theoretically \cite{P97, ZGN04, VP07, P10, ama7, ama9, IHA11, ama19, MGK13, S14, IMB15, BL15, IBH16, KBD17, IOR19, ama66, RIB22, ORE22, D22, ama85} and are now beginning to receive experimental verification \cite{XJS24}. Remarkably, even the motion of single objects is influenced by friction forces arising from their interaction with the fluctuations of the electromagnetic field \cite{EH1910, MMF02, KD02_2, ama7, ama9, GMK21, GMK22, SM22, GMK23}.  A paradigmatic example is the frictional force acting on an isotropic object moving at constant velocity relative to a thermal radiation bath, commonly referred to as thermal radiation drag \cite{MPP03, DK05, LDJ12, LDJ12_2, PH13, V15, MDL20}. 

The symmetry properties of the system play a crucial role in determining the characteristics and even the existence of Casimir interactions. For example, a Casimir torque only exists between objects with anisotropic response, which breaks the relevant rotational symmetries \cite{SGP18,ACG20}. Frictional phenomena, such as the thermal radiation drag, can also be understood as symmetry-breaking effects. In this case, the Doppler shift of the electromagnetic field in the frame comoving with the object causes an imbalance in the absorption of the field components propagating with and against it. Importantly, this effect is independent of the symmetry properties of the object. However, one may inquire what happens if, in addition, the object displays an anisotropic response, an aspect that has been overlooked and can give rise to significant effects.

\begin{figure}
\begin{center}
\includegraphics[width=80mm,angle=0]{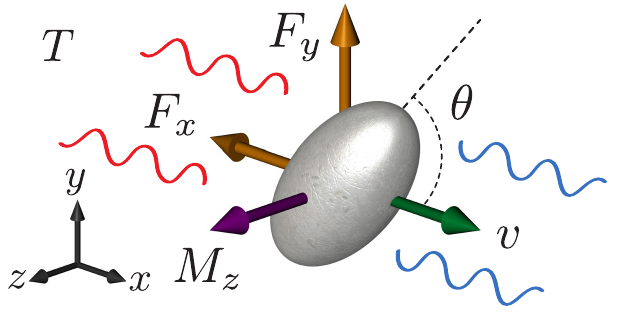}
\caption{ Schematic of a spheroidal nanostructure moving with velocity ${\bf v} = v \hat{\bf x}$ with respect to a thermal radiation bath at temperature $T$. The nanostructure has diameters $d_{\parallel}$ and $d_{\perp}$  along the directions parallel and perpendicular to the axis of revolution, which is contained in the $xy$-plane and forms an angle $\theta$ with the $x$-axis.} \label{fig1}
\end{center}
\end{figure}

In this Letter, we predict that nanostructures with anisotropic response moving with respect to a thermal radiation bath experience a force perpendicular to the direction of motion. This lateral force is linear in the velocity and, under the appropriate conditions, can be of the same order of magnitude as the drag force. In addition, the anisotropic nanostructure experiences a torque that depends quadratically on the velocity and tends to align one of its principal axes with the direction of motion\textemdash a behavior analogous to the reorientation of two anisotropic objects produced by the Casimir torque. 

In contrast to the previously studied drag force  \cite{MPP03, DK05, LDJ12, LDJ12_2, PH13, V15, MDL20}, the lateral force and torque predicted in this work require an anisotropic response. Indeed, while the drag force arises solely from the symmetry breaking in the thermal radiation bath induced by the motion of the nanostructure, the lateral force and torque originate from the interplay of that effect with the coupling between the electromagnetic field components polarized parallel and perpendicular to the direction of motion enabled by the anisotropy of the nanostructure. The results of this work advance our understanding of the nontrivial effects that the interaction with thermal radiation has on the motion of anisotropic nanostructures.
 
The system under study, which is depicted in Figure~\ref{fig1}, consists of a spheroidal nanostructure moving with constant velocity ${\bf v} = v \hat{\bf x}$ with respect to a thermal radiation bath at temperature $T$. The nanostructure is characterized by diameters $d_{\parallel}$ and $d_{\perp}$ along the directions parallel and perpendicular to the axis of revolution, respectively, which define an aspect ratio $\eta = d_{\parallel}/d_{\perp}$. As shown in Figure~\ref{fig1}, the axis of revolution is contained in the $xy$-plane and forms an angle $\theta$ with the $x$-axis. We assume that both $d_{\parallel}$ and $d_{\perp}$ are much smaller than the thermal wavelength $\lambda_{T} = 2\pi c \hbar / k_{\rm B}T$, which allows us to model its response as a point electric dipole $\bf p$ with a polarizability tensor 
\begin{equation}
{\boldsymbol{\alpha}}(\omega) = \begin{pmatrix}
\alpha_{xx} & \alpha_{xy} & 0\\
\alpha_{xy} & \alpha_{yy} & 0\\
0 & 0 & \alpha_{zz}
\end{pmatrix}.\nonumber
\end{equation}
This tensor encodes an anisotropic but reciprocal response and its components are $\alpha_{xx} = \alpha_{\parallel} \cos^2 \theta + \alpha_{\perp} \sin^2 \theta$, $\alpha_{yy} = \alpha_{\perp} \cos^2 \theta + \alpha_{\parallel} \sin^2 \theta$, $\alpha_{zz} = \alpha_{\perp}$, and $\alpha_{xy}=\alpha_{yx}=(\alpha_{\parallel}-\alpha_{\perp})\sin \theta \cos \theta$, with $\alpha_{\parallel}$ and $\alpha_{\perp}$ representing the polarizability along the directions parallel and perpendicular to the axis of revolution.

We choose to work in the reference frame comoving with the nanostructure since, as has been previously discussed \cite{V15}, the calculation in this frame provides a direct insight into the acceleration of the nanostructure (see the Appendix for details). The different components of the force are given by ${F_i} = \langle {\bf p}(t) \cdot [\partial_i {\bf E} ({\bf r}, t)]_{{\bf r} = {\bf 0}}\rangle$, where ${\bf E} ({\bf r}, t)$ is the electric field of the thermal bath, $\partial_i$ indicates the partial derivative with respect to the $i$th Cartesian component of $\bf r$, and $\langle \rangle$ denotes the average over fluctuations. Then, using the appropriate Fourier transforms, we get 
\begin{equation}
\begin{aligned}
F_i =&  -{\rm i} \int  \frac{{\rm d} \omega {\rm d} \omega^{\prime}}{(2 \pi)^2} \frac{{\rm d} {\bf k} {\rm d}{\bf k}^{\prime}}{(2 \pi)^6} {\rm e}^{-{\rm i} (\omega - \omega^{\prime})t} \\
& \times\!\! \sum_{j,l=x,y,z} \!\! k^{\prime}_i  \alpha_{jl}(\omega) \langle E^{\rm fl}_l({\bf k}, \omega) E_j^{{\rm fl}*}({\bf k^{\prime}}, \omega^{\prime}) \rangle, \label{eq_force}
\end{aligned}
\end{equation}
where $E^{\rm fl}_i(\mathbf{k},\omega)$ represents the different components of the fluctuating electric field. Notice that we do not consider the contribution of the fluctuating dipole because it vanishes in the reference frame comoving with the nanostructure \cite{V15}. The reason is that, in this frame, the electromagnetic field radiated by the nanostructure is identical in any two opposing directions, resulting in the cancellation of each component of the force. Furthermore, we ignore effects associated with retardation, which are also negligible in the limit of $d_{\parallel},d_{\perp}\ll \lambda_T$.

To calculate the average over the fluctuations of the electric field, we use the fluctuation-dissipation theorem, which, in the frame comoving with the nanostructure, reads  $\langle E^{\rm fl}_i({\bf k}, \omega) E^{\rm fl*}_j({\bf k^{\prime}}, \omega^{\prime}) \rangle = 32 \pi^4 \hbar \delta(\omega - \omega^{\prime})\delta({\bf k} - {\bf k}^{\prime}) {\rm Im}\{G_{ij}({\bf k},\omega)\} \left[n(\omega_{+})+1/2\right]$ \cite{PV1971, V15} (see the Appendix for details). In this expression
${\rm Im}\{G_{ij}({\bf k}, \omega)\}  =  (2 \pi^2/k) [\delta(k-\omega/c) - \delta(k+\omega/c)] [\delta_{ij} \omega^2/c^2 - k_i k_j]$ is the imaginary part of the Green's function of free space, $n(\omega) = [\exp(\hbar \omega / k_{\rm B} T)-1]^{-1}$ is the Bose-Einstein distribution, $k = |{\bf k}|$, $\omega_{+}=\gamma(\omega+v k_x)$ is the Doppler shifted frequency, $\gamma = 1 / \sqrt{1-\beta^2}$, and $\beta = v/c$.

For the integral of Eq.~\eqref{eq_force} to take nonzero values, the integrand must be even with respect to each Cartesian component of $\bf k$. Considering this and using the expression of the fluctuation-dissipation theorem given above, we obtain the thermal radiation drag force 
\begin{equation}
 F_x = \frac{\hbar c}{2 \pi^2} \int  \! {\rm d} {\bf k}  \frac{k_x}{k}  \!\!\sum_{i=x,y,z}  \!\! {\rm Im}\{ \alpha_{ii}(ck)\} (k^2-k_i^2) n(\Omega_+), \nonumber
\end{equation}
and the thermal radiation lateral force 
\begin{equation}
F_y = - \frac{\hbar c}{\pi^2}  \int   \! {\rm d} {\bf k} \frac{k_x k_y^2}{k} {\rm Im}\{\alpha_{xy}(ck)\}n(\Omega_+), \nonumber
\end{equation}
where $\Omega_+ = \gamma(ck+v k_x)$. These expressions are valid for arbitrary values of $v$. However, since generally $v\ll c$, we can approximate $n(\Omega_+)$ to the lowest order in $v$ as $n(\gamma ck + \gamma v k_x)  \approx n(ck) - \hbar v k_x /[4k_{\rm B} T \sinh^2(\hbar c k/ 2k_{\rm B} T)] $. Using this expression and computing the angular integrals, we obtain
\begin{equation}
F_x = -\frac{\beta \hbar^2}{15 \pi c^4 k_{\rm B}T} \int_0^{\infty} \!\! {\rm d}\omega \frac{\omega^5 {\rm Im}\{2{\rm Tr}\{\boldsymbol{\alpha}(\omega)\}-\alpha_{xx}(\omega) \}}{\sinh^2(\hbar \omega / 2k_{\rm B}T)},
\nonumber
\end{equation}
for the drag force and
\begin{equation}
F_y = \frac{\beta \hbar^2}{15 \pi c^4 k_{\rm B}T} \int_0^{\infty} \!\!  {\rm d}\omega \frac{\omega^5 {\rm Im}\{\alpha_{xy}(\omega)\}}{\sinh^2(\hbar \omega / 2k_{\rm B}T)},
\nonumber
\end{equation}
for the lateral force. 

When $\alpha_{xx}=\alpha_{yy}=\alpha_{zz}$, the expression for $F_{x}$ reduces to the thermal radiation drag for isotropic structures described in \cite{MPP03, PH13, V15, MDL20}. Regardless of the orientation of the nanostructure, $F_x$ always opposes its motion. However, since the trace of the polarizability tensor is invariant under rotations, $F_x$ can be minimized (maximized) by choosing the orientation for which $\alpha_{xx}$ is maximum (minimum).

In contrast, $F_y$ represents a lateral force that acts perpendicular to the motion of the nanostructure and, to the best of our knowledge, has not been previously reported. 
Its existence is linked to the presence of off-diagonal components in the polarizability tensor capable of mixing the directions parallel and perpendicular to the trajectory of the nanostructure. Furthermore, since the imaginary part of the polarizability tensor of a nanostructure with reciprocal response is symmetric and, hence, diagonalizable, there exists a spatial rotation that sets ${\rm Im}\{\alpha_{xy} \} = 0$. Therefore, the effect of $F_y$ is to alter the trajectory of the nanostructure until its velocity aligns with a set of axes for which its polarizability tensor is diagonal, thereby causing the lateral force to vanish.

\begin{figure}
\begin{center}
\includegraphics[width=80mm,angle=0]{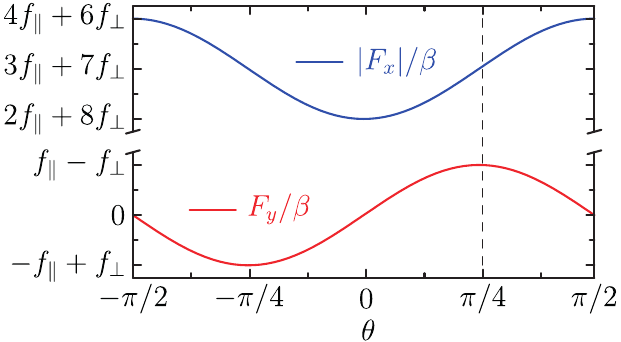}
\caption{Analysis of the dependence with $\theta$ of the thermal radiation forces experienced by a prolate ($\eta>1$) spheroidal nanostructure. } \label{fig2}
\end{center}
\end{figure}

We can further simplify the expressions for $F_x$ and $F_y$ to 
\begin{equation}
F_{x} = -(3f_{\parallel} + 7f_{\perp})\beta + (f_{\parallel} - f_{\perp}) \beta \cos 2\theta,\nonumber
\end{equation} 
and
\begin{equation}
F_{y} = (f_{\parallel} - f_{\perp}) \beta \sin 2\theta,\nonumber
\end{equation}
respectively, by introducing 
\begin{equation}
f_{\nu} =\frac{\hbar^2}{30 \pi c^4 k_{\rm B}T} \int_0^{\infty} \!\! {\rm d}\omega \frac{\omega^5 {\rm Im}\{\alpha_{\nu}(\omega) \}}{\sinh^2(\hbar \omega / 2k_{\rm B} T)},
\label{eq_fi}
\end{equation} 
where $\nu =\,\parallel, \perp$. These expressions explicitly show the dependence of the forces with $\theta$. Furthermore, since $f_{\nu}$ is a positive real number, $3f_{\parallel} + 7f_{\perp} > |f_{\parallel} - f_{\perp}|$, and hence, $F_x$ is always larger than $F_y$. However, as we demonstrate later, $F_y$ can be made to have the same order of magnitude as $F_x$ by appropriately choosing $\eta$. 

Figure~\ref{fig2} schematically analyzes the dependence of $F_x$ and $F_y$ with $\theta$ for a nanostructure with $\eta>1$, which translates into $f_{\parallel}>f_{\perp}$. As expected from the expressions above, $F_x$ reaches its maximum for $\theta = \pm \pi/2$, when the axis of revolution is perpendicular to the direction of motion. On the contrary, $F_y$ attains its extremal values at $\theta = \pm \pi/4$, for which the off-diagonal components of $\boldsymbol{\alpha}$ are maximized. Importantly, the sign of $F_y$ is determined by the sign of $\theta$. Moreover, $F_y$ vanishes for $\theta =0$ and $\theta = \pm \pi/2$. However, these two conditions are not equivalent. 
In particular, since $F_y$ acts in the direction perpendicular to the trajectory, $\theta = 0$ and $\theta = \pm \pi/2$ are the stable equilibrium conditions for $f_{\parallel}>f_{\perp}$ and $f_{\parallel}<f_{\perp}$, respectively. Therefore, the effect of the lateral force is to align the trajectory with the axis of the nanostructure along which its polarizability is maximum.

\begin{figure}
\begin{center}
\includegraphics[width=80mm,angle=0]{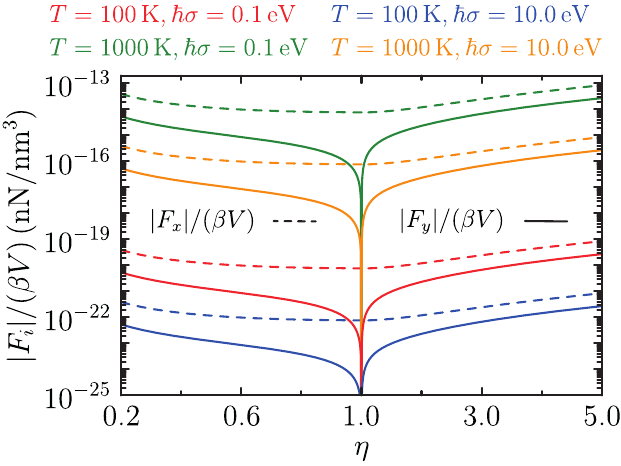}
\caption{Absolute value of the forces experienced by a spheroidal nanostructure as a function of $\eta$ for $\theta = \pi/4$. The dashed and solid curves represent the drag and lateral components of the force, respectively. } \label{fig3}
\end{center}
\end{figure}

To obtain a quantitative estimate of the thermal radiation drag and lateral forces, we particularize our results for a metallic nanostructure as a representative case of an absorbing system. We approximate its dielectric function using a Drude model as $\varepsilon \approx 1 + {\rm i} 4\pi \sigma / \omega$, which is valid at sufficiently low frequencies. Within this approximation, 
$\alpha_{\nu}(\omega) \approx V/(4\pi D_{\nu})+ {\rm i} V\omega/(16 \pi^2 \sigma D^2_{\nu})$ with $V=\pi d_{\parallel}d_{\perp}^2/6$ being the volume of the spheroidal nanostructure and $D_{\nu}$ the depolarization factor, which is determined by $\eta$ \cite{paper112}. Using these expressions, we can solve analytically the integral of Eq.~\eqref{eq_fi} to obtain
\begin{equation}
f_{\nu} = \frac{2\pi^3}{315}\frac{k^6_{\rm B}T^6}{c^4\hbar^5}  \frac{V}{\sigma D_{\nu}^2}. \nonumber
\end{equation}
Importantly, as shown in Figure~\ref{fig_S1}, this approximation provides an accurate estimate of the force for a wide range of temperatures. Figure~\ref{fig3} displays the absolute values of the two components of the thermal radiation force as a function of the aspect ratio $\eta$. We consider the configuration with $\theta = \pi/4$, for which $F_x = -(3f_{\parallel} + 7f_{\perp})\beta$ and $F_y = (f_{\parallel} - f_{\perp}) \beta$. We use dashed and solid curves to represent $F_x$ and $F_y$, respectively. 
Since $F_y$ originates in the anisotropy of the nanostructure, its magnitude strongly varies with $\eta$, following the dependence on $D_{\parallel}^{-2}-D_{\perp}^{-2}$. In the limit $\eta\gg1$, the nanostructure approaches a one-dimensional system with $f_{\parallel}\gg f_{\perp}$, which results in $|F_y/F_x| \approx 1/3$. Conversely, when $\eta \ll 1$, the nanostructure approaches a two-dimensional geometry with $f_{\parallel}\ll f_{\perp}$, which results in a smaller ratio $|F_y/F_x| \approx 1/7$. As expected, $F_y$ vanishes for $\eta=1$, when the nanostructure becomes a sphere. These results demonstrate that, by reducing the dimensionality of the nanostructure, it is possible to make the lateral force reach values of the same order of magnitude as the drag force. 
Interestingly, the value of $F_x$ also varies with $\eta$, significantly increasing as the anisotropy of the nanostructure grows.
Figure~\ref{fig3} also analyzes the dependence of $F_x$ and $F_y$ with $\sigma$ and $T$. Both components of the force increase with the sixth power of the temperature of the thermal radiation bath and decrease with the conductivity of the nanostructure. This is consistent with the dissipative origin of the thermal radiation forces, which benefit from a larger number of photons and increased absorption losses.

As discussed above, the lateral force alters the trajectory of the nanostructure, aligning it with the axis along which the polarizability is maximum. It seems natural, then, to inquire whether the nanostructure also experiences a torque that modifies its spatial orientation. Working, again, within the dipolar approximation, the torque acting on the nanostructure is given by ${\bf M} = \langle {\bf p}(t) \times {\bf E} ({\bf r}={\bf 0}, t)\rangle $.  As illustrated in Figure~\ref{fig1}, the only nonvanishing component of the torque acting on the spheroidal nanostructure under consideration is along the $z$-axis. Following the same steps as for the drag and lateral forces, we have
\begin{equation}
M_z = \frac{\hbar c}{2 \pi^2} \int \! {\rm d} {\bf k} \frac{k_x^2-k_y^2}{k} {\rm Re}\{\alpha_{xy}(ck)\}n(\Omega_+), \nonumber
\end{equation}
which, in the limit of small velocities $v\ll c$, after performing the angular integrals, becomes
\begin{equation}
M_z = \frac{\beta^2 \hbar^3}{30 \pi c^3 (k_{\rm B}T)^2} \int_0^{\infty} \!\!  {\rm d}\omega \, \omega^5 {\rm Re}\{\alpha_{xy}(\omega)\} \frac{\cosh\!\left(\frac{\hbar \omega}{2k_{\rm B}T}\right)}{\sinh^3\!\left(\frac{\hbar \omega}{2k_{\rm B}T}\right)}.
\nonumber
\end{equation}
Similar to the lateral force, the torque arises from the combined effect of the anisotropy of the thermal radiation bath and  the anisotropic optical response of the nanostructure. 
 As a result, both of them depend on $v$ and $\alpha_{xy}$. However, there are substantial differences between these two phenomena. While the lateral and drag forces are, to the lowest order, linear with respect to the velocity,  the torque exhibits a quadratic dependence on $v$, making it less significant at moderate velocities. Furthermore, the thermal radiation forces are proportional to the imaginary part of $\boldsymbol{\alpha}$, as expected from its dissipative character. Meanwhile, the torque depends on the real part of $\boldsymbol{\alpha}$, which is associated with a dispersive character. In that sense, the effect of the thermal radiation torque is similar to the conventional Casimir torque arising between two anisotropic media \cite{SGP18}, whose effect is to align their principal axes. The difference is that, in this case, the anisotropy of the thermal radiation bath plays the role of the second anisotropic medium. For that reason, unlike the conventional Casimir torque, the thermal radiation torque described here only exists at finite temperature.

\begin{figure}
\begin{center}
\includegraphics[width=80mm,angle=0]{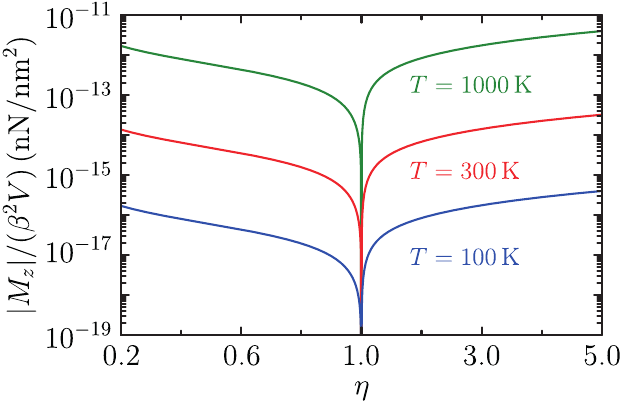}
\caption{Absolute value of the torque experienced by a spheroidal nanostructure as a function of $\eta$ for $\theta = \pi/4$.} \label{fig4}
\end{center}
\end{figure}

By introducing 
\begin{equation}
m_{\nu} =\frac{\hbar^3}{60 \pi c^3 (k_{\rm B}T)^2} \int_0^{\infty}\!\! {\rm d}\omega \, \omega^5 {\rm Re}\{\alpha_{\nu}(\omega)\}\frac{\cosh\!\left(\frac{\hbar \omega}{2k_{\rm B}T}\right)}{\sinh^3\!\left(\frac{\hbar \omega}{2k_{\rm B}T}\right)},\nonumber
\end{equation}
we can simplify the expression of $M_z$ to
\begin{equation}
M_z = (m_{\parallel} - m_{\perp}) \beta^2 \sin 2\theta. \nonumber
\end{equation}
This result explicitly shows the dependence of the torque with $\theta$, confirming that its effect is to align the axis of the nanostructure along which its polarizability is minimum with the direction of motion. This is in contrast with the effect of the lateral force, which, as discussed previously, tends to align the trajectory of the nanostructure with the axis of maximum polarizability. Therefore, the combined effect of the torque and the lateral force can give rise to interesting nontrivial dynamics.

Considering again a Drude dielectric function in the limit of low frequencies, we have
\begin{equation}
m_{\nu} = \frac{\pi^2}{45}\frac{k^4_{\rm B}T^4}{c^3\hbar^3} \frac{V}{D_{\nu}},
\nonumber
\end{equation} 
which allows us to analyze the numerical value of the torque for the spheroidal nanostructure under consideration. The corresponding results, normalized to $\beta^2$ and $V$, are shown in Figure~\ref{fig4} as a function of $\eta$. We take $\theta =\pi/4$ and consider three different values of $T$, as indicated by the legend. As the lateral force, the torque increases with both $\eta$ and $T$. However, its dependence with these parameters is less pronounced since it is proportional to $D_{\parallel}^{-1}-D_{\perp}^{-1}$ and $T^4$. 
Furthermore, the torque does not depend on the conductivity since, in this limit, the latter only contributes to dissipative processes.

In summary, we have predicted that an anisotropic nanostructure moving with respect to a thermal radiation bath experiences a lateral force and a torque. These phenomena arise from the interplay between the anisotropic response of the nanostructure and the asymmetry in the thermal radiation bath introduced by its motion. The effect of the lateral force is to align the trajectory of the nanostructure with the axis along which its polarizability is maximum. The torque, on the other hand, tends to rotate the nanostructure until its axis of minimum polarizability is parallel to the motion. Although we have focused on a spheroidal nanostructure, our results are general and apply to any system that can be modeled with an electric polarizability tensor. By considering a material response described with a Drude model, we have obtained analytical expressions for the lateral force and the torque. This has allowed us to understand the dependence of these phenomena on the characteristics of the nanostructure  such as its volume, aspect ratio, and conductivity, as well as the temperature of the thermal bath. In doing so, we have found that the lateral force can be of the same order as the drag force for nanostructures with reduced dimensionality. While here we work with a nanostructure made of an isotropic material and introduce anisotropy through its geometry, similar results can also be obtained with a sphere made of an anisotropic material, as explored in Figure~\ref{fig_S2}.
 
Our results can be seen as the thermal radiation analog of an object moving in a viscous fluid. When an anisotropic object moves in a viscous fluid with the appropriate orientation, in addition to a drag force, the interaction with the fluid can produce a lateral force and a torque \cite{A1990}. The paradigmatic example is a wing, whose geometry is designed to achieve a lateral force strong enough to compensate gravity. In our work, the thermal radiation bath plays the role of the viscous fluid, while the anisotropy of the object and its orientation come into play through the polarizability tensor. Our work sheds light on the complex motion of anisotropic nanostructures moving in a thermal radiation bath and extends the range of known electromagnetic analogs to fluid dynamics phenomena, such as the electromagnetic Magnus effect \cite{H00_2,M22}.   Furthermore, we envision that analogs to our results could be explored utilizing engineered fluctuating fields \cite{PM18}, whose intensity can be controlled externally, allowing for forces and torques only limited by the available power. An alternative could be to employ approaches based on synthetic motion in the context of time-varying media \cite{HVR24}.

\begin{acknowledgements}
The authors acknowledge support from Grant No.~PID2022-137569NB-C42 funded by MICIU/AEI/10.13039/501100011033 and FEDER, EU.  J.R.D-R. acknowledges support from a predoctoral fellowship from the MCIN/AEI assigned to Grant No. PID2019-109502GA-I00. 
\end{acknowledgements}

\onecolumngrid
\appendix
\section{Appendix}\label{ap}
%\renewcommand{\thefigure}{S\arabic{figure}}
%\etcounter{figure}{0} 

\subsection{Relation between the force and the acceleration in the different reference frames}

As in the main text, we consider a nanostructure moving with constant velocity ${\bf v} = v {\bf \hat{x}}$.  We denote the variables in the reference frame comoving with the nanostructure without any special markings and add a tilde $\sim$ to denote the corresponding variables in the laboratory frame. To calculate the relation between the force and the acceleration in the reference frame comoving with the nanostructure, we start from the definition of linear momentum \cite{J99}:
\begin{equation}
{\bf p} = \frac{m {\bf u}}{ \sqrt{1 - u^2/c^2}}, \nonumber
\end{equation}
where $m$ is the rest mass of the nanostructure and ${\bf u}$ its velocity. The force can be written as 
\begin{equation}
{\bf F} = \frac{{\rm d} {\bf p}}{{\rm d} t} = \left(\frac{{\rm d}m}{{\rm d}t}\right) \frac{{\bf u}}{ \sqrt{1 - u^2/c^2}} + m \left(\frac{{\rm d} {\bf u}}{{\rm d}t}\right) \frac{1}{ \sqrt{1 - u^2/c^2}} + m {\bf u} \frac{{\rm d}}{{\rm d}t} \left(\frac{1}{\sqrt{1 - u^2/c^2}}\right). \nonumber
\end{equation}
Since the velocity of the nanostructure in the reference frame comoving with it tends to zero, taking the limit $u \rightarrow 0 $, the expression above reduces to
${\bf F} = m {\bf a}$, where ${\bf a} = {\rm d} {\bf u}/{\rm d} t$ is the acceleration of the particle. This equation shows that second Newton's law is fulfilled in the frame comoving with the nanostructure, providing a direct physical interpretation of the force. 
  
The components of the acceleration in the laboratory frame, ${\bf \tilde{a}}$, can be expressed in terms of the components of the acceleration in the frame comoving with the nanostructure, ${\bf a}$, by performing time derivatives of the Lorentz transformations for the velocity
\begin{align}
\tilde{u}_x={}&\frac{u_x+v}{1+v u_x / c^2},  \nonumber \\
\tilde{u}_y={}&\frac{u_y}{\gamma (1+v u_x / c^2)}, \nonumber \\
\tilde{u}_z={}&\frac{u_z}{\gamma (1+v u_x / c^2)}. \nonumber 
\end{align}
By doing so, and taking the limit $\tilde{u} \rightarrow v $ and $ u \rightarrow 0 $, we obtain after some algebra $\tilde{a}_x = a_x / \gamma^3$ and $\tilde{a}_y = a_y / \gamma^2$. Therefore, if we only consider terms that are linear on $v/c$, as we did in the main text, we have that  $\tilde{\bf a} = {\bf a}$. This means that the force in the frame comoving with the nanostructure is also proportional to the acceleration in the laboratory frame ${\bf F} = m \tilde{\bf a}$.

\subsection{Fluctuation-dissipation theorem in reciprocal space}

In the laboratory reference frame, where the thermal radiation bath is at rest, the fluctuation-dissipation theorem for the electric field is given by \cite{ama7}
\begin{equation}
\langle \tilde{E}^{\rm fl}_i({\bf \tilde{r}}, \tilde{\omega}) \tilde{E}_j^{{\rm fl}*}({\bf \tilde{r}}^{\prime}, \tilde{\omega}^{\prime}) \rangle=4  \pi \hbar \delta(\tilde{\omega}-\tilde{\omega}^{\prime}){\rm Im}\{G_{ij}({\bf \tilde{r}} - {\bf \tilde{r}^{\prime}} , \tilde{\omega})\}\left[n(\tilde{\omega})+\frac{1}{2}\right],
\nonumber
\end{equation}
where $n(\omega) = [\exp(\hbar \omega / k_{\rm B} T)-1]^{-1}$ is the Bose-Einstein distribution and $G_{ij}({\bf \tilde{r}} , \tilde{\omega}) = (\delta_{ij} \tilde{\omega}^2/c^2+ \partial_i \partial_j) {\rm e}^{{\rm i}\tilde{\omega}\tilde{r}/c}/\tilde{r}$ is the Green's function of free space.
By using the Fourier transform of the fluctuating electric field, defined as
\begin{equation}
\tilde{E}^{\rm fl}_i({\bf \tilde{k}},\tilde{\omega}) = \int \!\! {{\rm d}{\bf r}}  \tilde{E}^{\rm fl}_i({\bf \tilde{r}}, \tilde{\omega}) {\rm e}^{-{\rm i}{\bf \tilde{k}}\cdot {\bf \tilde{r}}} ,
\nonumber
\end{equation} 
we obtain
\begin{equation}
\langle \tilde{E}^{\rm fl}_i({\bf \tilde{k}}, \tilde{\omega}) \tilde{E}_j^{{\rm fl}*}({\bf \tilde{k}}^{\prime}, \tilde{\omega}^{\prime} )\rangle  = 32 \pi^4 \hbar \delta(\tilde{\omega}-\tilde{\omega}^{\prime}) \delta({\bf \tilde{k}} - {\bf \tilde{k}^{\prime}}) \left[n(\tilde{\omega})+\frac{1}{2}\right] {\rm Im}\{G_{ij}({\bf \tilde{k}}, \tilde{\omega}) \},
\nonumber
\end{equation} 
where
\begin{equation}
{\rm Im}\{G_{ij}({\bf \tilde{k}}, \tilde{\omega})\} = \frac{2\pi^2}{\tilde{k}}\left[\delta(\tilde{k}-\tilde{\omega}/c) - \delta(\tilde{k}+\tilde{\omega}/c) \right] \left[\delta_{ij} \tilde{\omega}^2/c^2 - \tilde{k}_i \tilde{k}_j\right]  .
\nonumber 
\end{equation} 
is the imaginary part of the Green's function of free space expressed in reciprocal space \cite{LP1981}. Therefore, the fluctuation-dissipation theorem in reciprocal space reads
\begin{equation}
\begin{aligned}
\langle \tilde{E}^{\rm fl}_i({\bf \tilde{k}}, \tilde{\omega}) \tilde{E}_j^{{\rm fl}*}({\bf \tilde{k}}^{\prime}, \tilde{\omega}^{\prime} )\rangle = (2\pi)^6 \hbar \delta(\tilde{\omega}-\tilde{\omega}^{\prime}) \delta({\bf \tilde{k}} - {\bf \tilde{k}^{\prime}}) \left[\delta(\tilde{k}-\tilde{\omega}/c) - \delta(\tilde{k}+\tilde{\omega}/c)\right]\frac{1}{\tilde{k}} \left[\delta_{ij}\tilde{\omega}^2/c^2 - \tilde{k}_i \tilde{k}_j\right] \left[n(\tilde{\omega})+\frac{1}{2}\right].
\end{aligned}
\label{eq_S1}
\end{equation} 
 
\subsection{Transformation of the fluctuation-dissipation theorem to the moving frame}

Throughout the main text, we choose to work in the reference frame comoving with the nanostructure. However, since the fluctuation-dissipation theorem derived in the previous section is formulated in the laboratory frame, we need to find the appropriate transformation for Eq.~\eqref{eq_S1}. To that end, we start from the Lorentz transformation for the electric field \cite{J99}
\begin{subequations}
\begin{align}
E_x & =\tilde{E}_x, \label{eq_S2a} \\
E_y & = \gamma[\tilde{E}_y - \beta \tilde{H}_z] = \gamma\left[\tilde{E}_y-\beta \frac{\tilde{k}_x}{\tilde{k}} \tilde{E}_y+\beta \frac{\tilde{k}_y}{\tilde{k}} \tilde{E}_x\right],  \label{eq_S2b} \\
E_z & =\gamma[\tilde{E}_z + \beta \tilde{H}_y] = \gamma\left[\tilde{E}_z+\beta \frac{\tilde{k}_z}{k} \tilde{E}_x-\beta \frac{\tilde{k}_x}{\tilde{k}} \tilde{E}_z\right],  \label{eq_S2c}
\end{align}
\end{subequations}
where $\tilde{H}_i$ are the components of the magnetic field in the laboratory frame that satisfy ${\bf \tilde{H}}=({\bf \tilde{k}}\times{\bf \tilde{E}})/\tilde{k}$. Note that, for readability, we omit the dependencies on frequency and wavevector. We also need to account for the transformation of $\bf \tilde{k}$, as well as the Doppler shift in $\tilde{\omega}$, given by
\begin{subequations}
\begin{align}
& \tilde{\omega}=\gamma\left[\omega +\beta c k_x\right],  \label{eq_S3a}\\
& \tilde{k}_x=\gamma\left[k_x+\beta \omega/c\right], \label{eq_S3b} \\
& \tilde{k}_y=k_y, \label{eq_S3c} \\
& \tilde{k}_z=k_z. \label{eq_S3d}
\end{align}
\end{subequations}

To compute the average over fluctuations $\langle E^{\rm fl}_i({\bf k},\omega) E_j^{{\rm fl}*}({\bf k}^{\prime},\omega^{\prime}) \rangle$, we transform the fields using Eqs.~\eqref{eq_S2a}-\eqref{eq_S2c} and apply the fluctuation-dissipation theorem given in Eq.~\eqref{eq_S1}. Then, using Eqs.~\eqref{eq_S3a}-\eqref{eq_S3d},  we obtain an expression that depends on the components of the wavevector as $\langle E^{\rm fl}_i({\bf k},\omega) E_j^{{\rm fl}*}({\bf k}^{\prime},\omega^{\prime}) \rangle \propto  [k^2 \delta_{ij} - k_i k_j]$, which has the exact same form as in the laboratory frame. 

The direct application of Eqs.~\eqref{eq_S3a}-\eqref{eq_S3d} also shows that
\begin{equation}
\delta(\tilde{\omega}-\tilde{\omega}^{\prime}) \delta(\tilde{{\bf k}} - {\tilde{{\bf k}}^{\prime}}) = \delta(\omega-\omega^{\prime}) \delta({\bf k} - {\bf k^{\prime}}). \nonumber
\end{equation}
Similarly, since $\delta(x^2-a^2) = [\delta(x-a) + \delta(x+a)]/2|a|$ and $\tilde{k}^2-\tilde{\omega}^2/c^2 = k^2-\omega^2/c^2$, we have that 
\begin{equation}
 [\delta(\tilde{k}-\tilde{\omega}/c) - \delta(\tilde{k}+\tilde{\omega}/c)]\frac{1}{\tilde{k}} = [\delta(k-\omega/c) - \delta(k+\omega/c)]\frac{1}{k}, \nonumber
\end{equation}
while the Bose-Einstein distribution transforms as $n(\tilde{\omega}) = n(\gamma \omega+\gamma vk_x)$.

Therefore, combining all of these transformations together, the fluctuation-dissipation theorem in the frame comoving with the nanostructure is given by
\begin{equation}
\langle E^{\rm fl}_i({\bf k}, \omega) E_j^{{\rm fl}*}({\bf k}^{\prime}, \omega^{\prime} )\rangle = (2\pi)^6 \hbar \delta(\omega-\omega^{\prime}) \delta({\bf k} - {\bf k^{\prime}}) \left[\delta(k-\omega/c) - \delta(k+\omega/c)\right]\frac{1}{k} \left[\delta_{ij}\omega^2/c^2 - k_i k_j\right] \left[n(\gamma \omega + \gamma vk_x)+\frac{1}{2}\right].
\nonumber
\end{equation}

\setcounter{figure}{0}
\renewcommand{\figurename}{FIG.}
\renewcommand{\thefigure}{S\arabic{figure}}

\begin{figure}[h!]
\begin{center}
\includegraphics[width=180mm,angle=0]{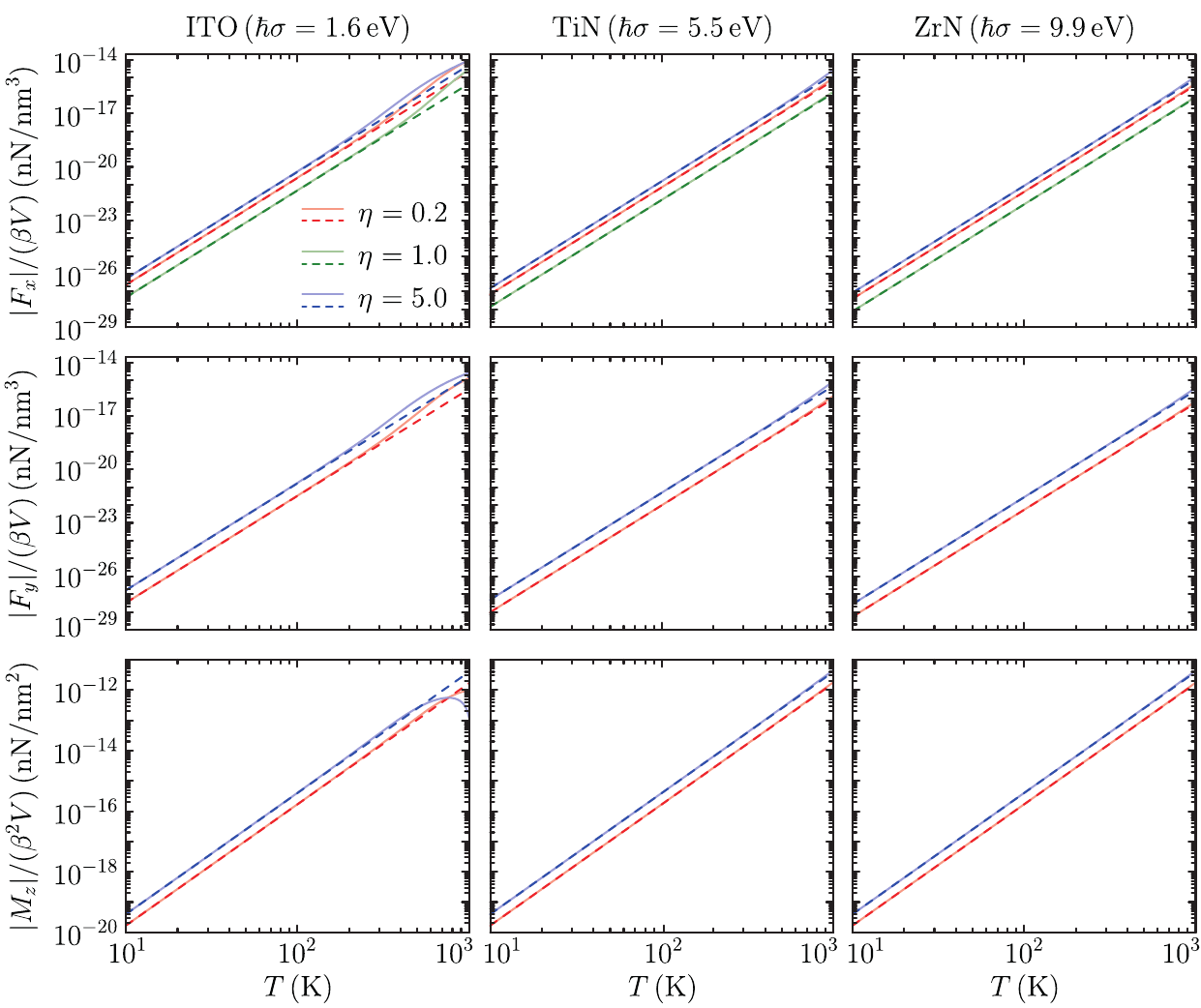}
\caption{Absolute value of the components of the force and the torque experienced by a spheroidal nanostructure made of different realistic materials as a function of $T$, for $\theta = \pi/4$. The top, middle, and bottom rows display, respectively, the drag force, the lateral force, and the torque, while the left, center, and right columns show the results for ITO, TiN, and ZrN. In all of the panels, the dashed curves represent the low-frequency approximation described in the main text, while the solid curves correspond to fully numerical calculations. Each color set denotes a different aspect ratio, as indicated by the legends. The fully numerical results are obtained by evaluating the frequency integrals appearing in the definitions of $f_{\nu}$ and $m_{\nu}$ using a Drude-Lorentz dielectric function $\varepsilon = \varepsilon_{\rm b} -\omega_{\rm p}^2/(\omega^2+{\rm i}\omega\gamma_{\rm p})+f_1\omega_1^2/(\omega_1^2-\omega^2-{\rm i}\omega\gamma_1)$ with parameters taken from Table 2 of Ref.~\cite{NSB13}. For the low-frequency approximation, we use the analytical expressions derived in the main text with a conductivity computed as $\sigma = \omega_{\rm p}^2 / (4 \pi \gamma_{\rm p})$, which yields $\hbar\sigma = 1.6$, $5.5$, and $9.9\,$eV for ITO, TiN, and ZrN, respectively. 
Comparing the two approaches, we observe that, for TiN and ZrN, which are the materials with the highest values of $\gamma_{\rm p}$ and $\sigma$, the low-frequency approximation agrees perfectly with the numerical calculations across the entire range of temperatures under consideration. In contrast, for ITO, the low-frequency approximation underestimates the value of the force for temperatures $T>300\,$K and overestimates the torque for $T>600\,$K.} \label{fig_S1}
\end{center}
\end{figure}

\begin{figure}
\begin{center}
\includegraphics[width=180mm,angle=0]{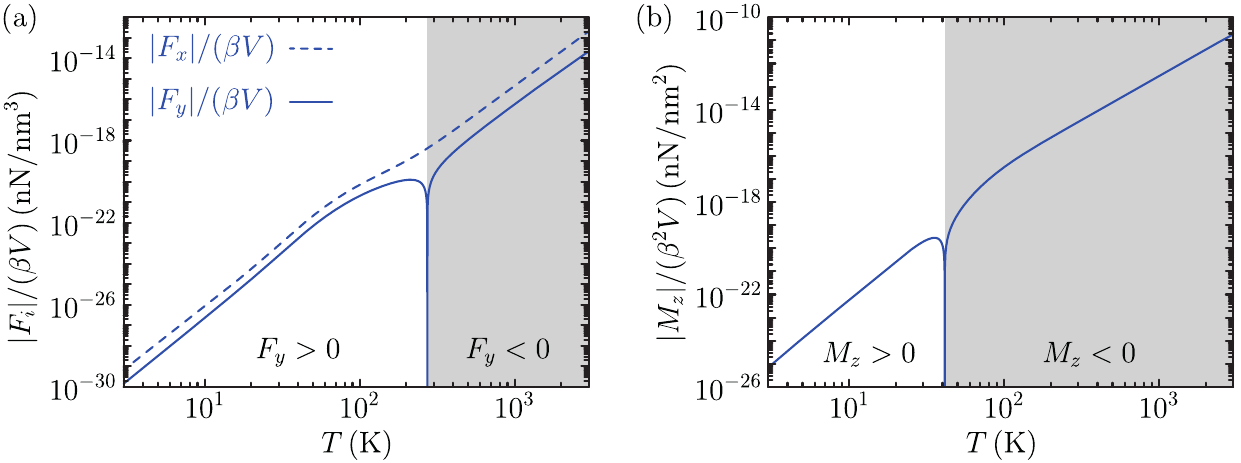}
\caption{Absolute value of the components of the force (a) and the torque (b) experienced by a spherical nanostructure of radius $R=10\,$nm made of graphite as a function of $T$. The grey background indicates the range of temperatures for which the lateral force and the torque are negative. All of the results are obtained by evaluating the frequency integrals appearing in the definitions of $f_{\nu}$ and $m_{\nu}$ using the dielectric funtion for graphite tabulated in Ref.~\cite{D03}, extended to lower frequencies using a Drude model. Since graphite is an uniaxial crystal, characterized by permitivity components $\varepsilon_{\parallel}$ along its optic axis and $\varepsilon_{\perp}$ along the orthogonal directions, we compute the parallel and perpendicular components of the polarizability of the graphite nanosphere using $\alpha_{\nu} = R^3(\varepsilon_{\nu} - 1) / (\varepsilon_{\nu} + 2)$ \cite{AGF10},  assuming that the optic axis lies in the $xy$-plane and forms an angle $\theta=\pi/4$ with the $x$-axis. Upon examining the results, we observe that, similar to the spheroidal nanostructure analyzed in the main text, both the force and the torque exhibit a strong temperature dependence. However, a key difference is that, in this case, the lateral force and torque can be either positive or negative, depending on the dominant component of the dielectric tensor in the relevant frequency range. } \label{fig_S2}
\end{center}
\end{figure}

%\bibliographystyle{apsrev}
%\bibliography{../../../refs}


\end{document}